\documentclass{iaus}
\usepackage{graphicx}

\title[The Nature and Nurture of Star Clusters] 
{The Nature and Nurture of Star Clusters}

\author[Bruce G. Elmegreen]   
{Bruce G. Elmegreen}

\affiliation{IBM Research Division, T.J. Watson Research Center\\
1101 Kitchawan Road, Yorktown Hts., NY 10598 USA, bge@us.ibm.com}

\pubyear{2010}
\volume{266}  
\pagerange{119--126}
 \jname{Star Clusters Basic Galactic Building
Blocks Throughout Time And Space} \editors{eds. Richard de Grijs and
Jacques Lepine}
\begin{document}

\maketitle

\begin{abstract}
Star clusters have hierarchical patterns in space and time, suggesting
formation processes in the densest regions of a turbulent interstellar
medium.  Clusters also have hierarchical substructure when they are
young, which makes them all look like the inner mixed parts of a
pervasive stellar hierarchy. Young field stars share this distribution,
presumably because some of them came from dissolved clusters and others
formed in a dispersed fashion in the same gas. The fraction of star
formation that ends up in clusters is apparently not constant, but may
increase with interstellar pressure. Hierarchical structure explains
why stars form in clusters and why many of these clusters are
self-bound. It also explains the cluster mass function. Halo globular
clusters share many properties of disk clusters, including what appears
to be an upper cluster cutoff mass. However, halo globulars are
self-enriched and often connected with dwarf galaxy streams. The mass
function of halo globulars could have initially been like the power law
mass function of disk clusters, but the halo globulars have lost their
low mass members. The reasons for this loss are not understood. It
could have happened slowly over time as a result of cluster
evaporation, or it could have happened early after cluster formation as
a result of gas loss. The latter model explains best the observation
that the globular cluster mass function has no radial gradient in
galaxies. \keywords{open clusters and associations, Galaxy: solar
neighborhood, galaxies: star clusters, stars: formation}
\end{abstract}

\firstsection 
\section{Introduction}
Star formation is hierarchical in time and space, suggesting turbulent
processes partition the gas. Star formation is also clustered to
varying degrees, suggesting the action of some dimensionless quantity
such as the turbulent Mach number. Here we review recent observations
of clusters and associations with an emphasis on hierarchical structure
for the young regions, the resulting mass functions for open and
globular clusters, and the likely dwarf galaxy origin for some halo
globular clusters in the Milky Way.

\section{Hierarchical Structure in Young Clusters}
\subsection{Spatial Correlations}

Piskunov et al. (2006) identified several ``open cluster complexes''
according to groupings of position, velocity, and age of clusters in
the solar neighborhood (see also Lynga 1982; Kharchenko et al. 2005).
Examples of these complexes included one of intermediate age consisting
of 7 clusters in Perseus-Auriga, an older one with 6 clusters in
Hyades, and a younger one with 23 members that is essentially Gould's
Belt. A typical complex spans a distance of several hundred parsecs.
They are probably the same objects that have been called ``star
complexes'' by Efremov (1995), who found them as groupings of Cepheid
variables and red supergiants. Ivanov (2005) catalogued star complexes
in M33 based on clusterings of blue stars, HII regions and WR stars.

de la Fuente Marcos \& de la Fuente Marcos (2008) identified five Open
Cluster Complexes from cluster positions and velocities within 2.5 kpc
of Sun: at a galactic longitude of $l=12^\circ$ and a distance of 1300
pc there is the Scutum-Sagittarius complex; at $l=75^\circ$ and 1400
pc, the Cygnus complex; at $l=132^\circ$ and  2000 pc, the
Cassiopeia-Perseus complex; at $l=200^\circ$ and 500 pc, the Orion
complex, and at $l=295^\circ$ 2000 pc, the Centaurus-Carina complex.
They suggest that open cluster complexes are fragments from a common
gas cloud. This is the usual explanation for star complexes too.

The gas clouds that form star complexes contain $\sim10^7\;M_\odot$ of
HI or H$_2$, and are fragmented into giant molecular clouds and
molecular cloud cores that form OB associations and OB subgroups or
clusters, respectively (e.g., Elmegreen 2007). These enormous regions
usually occur in galactic spiral arms with a separation of several
kiloparsecs (McGee \& Milton 1964; Elmegreen \& Elmegreen 1987).
Presumably the largest clouds are formed by galactic-scale processes
(i.e., their mass is the Jeans mass of the ambient ISM), and then
fragment into GMCs, GMC clumps, and eventually star-forming cores. Most
GMCs in the Milky Way (Grabelski et al. 1987) and M33 (Engargiola et
al. 2003) are in these gas giants, which are primarily HI. In M51,
where the gas is more molecular overall because of the higher pressure,
the gas giants are mostly molecular (Rand \& Kulkarni 1990).

Not all star formation produces dense clusters. Elias, Alfaro \&
Cabrera-Cano (2009) showed that clustering varies spatially in Gould's
Belt, with a larger fraction of stars ending up in clusters in the
Orion region than in the Sco-Cen region. This gradient in cluster
fraction corresponds to a gradient in young cluster density, suggesting
that the dense, high-pressure centers of star complexes produce stars
more efficiently than low-pressure peripheral regions. Higher
efficiency means that the final ratio of star-to-gas mass is higher,
and then the region is more likely to end up self-bound as a cluster
after the gas leaves.

Hierarchical structure extends from star complexes to embedded clusters
to individual young stars inside embedded clusters.  Feitzinger et al.
(1984, 1987) were among the first to recognize hierarchical or fractal
structure in large-scale star-forming regions.  An important point
about hierarchical structure is that the average density of gas
increases down the hierarchy, i.e., toward smaller fragment masses. The
mass fraction represented by dense star-forming clumps increases along
this sequence too. Bound regions require a high mass fraction for stars
and so appear only at the bottom of the hierarchy. This is the primary
reason why bound clusters are dense (much denser than the background
tidal limit, which is all they would need for self-boundedness at zero
velocity dispersion). With hierarchical structure, the average gas
density is high in regions where the star formation efficiency is high.
This is the reason why most star formation begins in the form of a
dense embedded cluster (Elmegreen 2008).

Scheepmaker et al. (2009) studied clusters in M51. Autocorrelation
functions for these clusters in three age bins show the youngest sample
is well correlated: it is hierarchical with a fractal dimension of
$\sim1.6$. The autocorrelation means that clusters are inside cluster
pairs and triplets, and these are inside clusters complexes and so on
up to scales greater than 1 kpc.  Clusters in the Antennae galaxy are
also auto-correlated up to $\sim1$ kpc (Zhang, Fall \& Whitmore 2001).
S\'anchez \& Alfaro (2008) studied HII regions in a number of galaxies.
They found that the positions of the brightest HII region have the
smallest fractal dimensions, which means they are still hierarchical.
For the galaxy NGC 6946, the high brightness HII regions have a fractal
dimension $D_c = 1.64$, those with medium brightness have $D_c = 1.82$,
and those with low brightness have $D_c = 1.79$. S\'anchez \& Alfaro
(2008) also found that $D_c$ decreases slightly with galaxy brightness,
which means that low-luminosity galaxies have more hierarchical
structure in the positions of their HII regions. Presumably these low
luminosity galaxies have less shear to smooth out the hierarchical
birth positions of the HII regions.

Elmegreen et al. (2006) studied the size distribution of star-forming
regions in NGC 628 by box-counting, which is a common technique for
measuring fractal dimensions. They used a high resolution image of this
galaxy from the ACS instrument on HST, and blurred it in successive
stages. The number of regions at each blurring stage was counted with
SExtractor. They found that the cumulative size distribution of
star-forming regions is $n(R)dR\sim R^{-2.5}dR$ for size $R$, and they
fit this to a projected fractal Brownian motion density distribution
having a 3D power spectrum with the Kolmogorov slope, $-3.66$.
Elmegreen et al. (2003) measured azimuthal intensity profiles of
optical light from whole galaxies and found that they have power-law
power spectra like turbulence too. For example, the young stars and
dust clouds seen in optical images of the flocculent galaxy NGC 5055
have the same scale-free distribution as the HI gas in the LMC. Both of
them give a Kolmogorov power spectrum. Block et al. (2009) presented
power spectra of Spitzer Space Telescope images of several galaxies,
using the near-IR passbands. They found that the passbands dominated by
stars (channels 1 and 2) had power spectra from noise plus the point
spread function that comes from the unresolved stellar images, whereas
the passbands dominated by dust and PAH emission (channel 4) had
Kolmogorov power spectra, clearly different in slope from the
star-dominated power spectra. Images of the range of Fourier components
that give the power law power spectrum in channel 4 show the resolved
hierarchical part of the galaxy.

\subsection{Time Correlations}

Clusters are also correlated in time in the sense that the age
difference increases with separation. Efremov \& Elmegreen (1998) found
this for $\sim600$ clusters in the LMC, and de la Fuente Marcos \& de
la Fuente Marcos (2009a) found it for local Milky Way clusters. The
slope of the correlation is about 0.5. This is interpreted as an
indication that the duration of star formation in a complex increases
as the square root of the size (Elmegreen \& Efremov 1996). Such a
relation between time and size also applies to self-gravitating clouds
with a common pressure: self-gravity means that $GM/R\sim5 V^2$, and a
common pressure $P$ means that $0.1GM^2/R^4\sim P$ (for mass $M$,
radius $R$, and velocity dispersion $V$). Eliminating $M$ gives
$V^2/R=(0.4PG)^{0.2}$, which is constant. As a result, the crossing
time is $R/V=R^{0.5}\left(0.4PG\right)^{-0.25}$. This is the same form
as the observed relation between duration of star formation and region
size. For typical $P=10^6k_{\rm B}$, $R/V=2.3\times10^{4}R^{0.5}$ in
cgs units, or $R/V=1.3R^{0.5}$ Myr for $R$ in pc and $V$ in km
s$^{-1}$. Efremov \& Elmegreen (1998) found for an age range of 1-100
Myr in the LMC that the time separation equals 30 Myr times the spatial
separation in degrees to the power 0.33 (for the LMC, 1 degree $=960$
pc). This is about the same relation as in the above simple derivation.
For the Milky Way, de la Fuente Marcos \& de la Fuente Marcos (2009a)
found a time separation equal to 11.1 Myr times the spatial separation
in parsecs to the power 0.16. Both the LMC and Milky Way have good
correlations, but the slopes are different.  de la Fuente Marcos \& de
la Fuente Marcos (2009a) suggest that if cluster disruption is taken
into account, the slopes are a bit steeper, between $\sim0.3$ for the
Milky Way and $\sim0.5$ for the LMC. If clusters within 2 kpc of the
Sun are considered, then the local slope could be $\sim0.4$.

\subsection{Cluster Pairs}

The time-distance correlation among young clusters implies that there
should be pairs of clusters born at about the same time and place.
Indeed, such pairs are well observed.  They were originally discovered
in the LMC by Bhatia \& Hatzidimitriou (1988) and in the SMC by
Hatzidimitriou \& Bhatia (1990). Dieball, Muller, \& Grebel (2002)
studied them again in the LMC.  de la Fuente Marcos \& de la Fuente
Marcos (2009b) studied cluster pairs in the Milky way and assessed
whether they are interacting. For example, they found that NGC 3293 and
NGC 3324 near $\eta$ Carinae are weakly interacting; their age
difference is 4.7 Myr. Similarly, NGC 659 and NGC 663 are weakly
interacting and their age difference is 19.1 Myr. Cluster pairs are the
closest members of the hierarchy of clusters found by autocorrelation
studies, power spectra, fractal analysis, and box-counting techniques.

\subsection{Individual Stars}

Individual young stars are correlated even if they are not in clusters.
The two point correlation function for stars in Taurus has a power law
form (Gomez et al. 1993; Larson 1995). Bastian et al. (2009) looked at
the positions and ages of stars in the LMC. They included 2000 sources
in each of several age intervals in the color-magnitude diagram. The
two point correlation function for these field stars has a slope that
gradually changes from $-0.5$ at the youngest age to 0 at an age of 175
Myr.  The coefficient in the 2-point correlation function goes
systematically to zero along this age sequence too. They also studied
the Cartwright \& Whitworth (2004) $Q$ parameter, which is the ratio of
the average minimum spanning tree length to the average correlation
length.  When a distribution of objects is fragmented into pieces, $Q$
is less than 0.8 or so; when it is smooth or with a smooth gradient,
$Q$ is closer to 1. Bastian et al. (2009) found that $Q$ also has a
gradual variation with stellar age, from $\sim0.55$ to $\sim0.75$ as
age increases to 175 Myr. This is consistent with the trend in the
2-point correlation function: older field stars are less clumped.
Gieles, Bastian, \& Ercolano (2008) did the same analysis for SMC
stars, finding the same age trends in the 2-point correlation function
and $Q$ out to about 100 Myr.

Odekon (2006) determined a ``correlation dimension,'' $d_c$, for stars
seen by HST in dwarf galaxies. The correlation dimension comes from the
equation $N(r)\propto r^{d_c}$ where $N(r)$ is the average number of
stars in a region of radius $r$. She found that the brightest dwarf has
the largest $d_c$ and the faintest dwarf has the smallest $d_c$. This
is consistent with the trend for HII region clustering in galaxies
found by S\'anchez \& Alfaro (2008).

\subsection{Internal Cluster Structure}

The hierarchy of young stellar structure often persists even inside
currently-forming clusters. Testi et al. (2000) found substructure in
an embedded IR cluster in Serpens, Smith et al. (2005) found it in the
$\rho$ Oph region, and Dahm \& Simon (2005) found four subclusters with
$\sim$Myr age differences in NGC 2264. Azimuthal profiles of young
clusters have non-Poisson distributions as well (Gutermuth et al.
2005). A recent X-ray map of young stars in NGC 6334 (Feigelson et al.
2009) shows substantial sub-structure in the positions of soft and hard
X-ray sources (which correspond to embedded populations with less than
and greater than 10 magnitudes of visual extinction, respectively). The
X-ray maps are nearly complete for stars more massive than the Sun.
S\'anchez \& Alfaro (2009) show that for 16 Milky Way clusters, the
stars in the younger, bigger clusters are more clumped. Evidently, as a
cluster ages, its subclusters mix and the substructure smooths out.

Schmeja, Kumar, \& Ferreira (2008) studied clumpy structure in four
embedded clusters.  For IC 348, NGC 1333, and the $\rho$ Oph region,
$Q$ is lower (more clumpy) for class 0/1 objects (young) than for class
2/3 (old). Also, among four subclumps in $\rho$ Oph, $Q$ is lower and
it is more gaseous where class 0/1 objects dominate, and $Q$ is also
lower for class 0/1 alone than for class 2/3. For an even younger
stage, Johnstone et al. (2000, 2001), Enoch et al. (2006), and Young et
al. (2006) found that the mm-wave pre-stellar clumps in several regions
are spatially correlated.

\subsection{Summmary: "Clusters" are the Dense Cores of a Pervasive
Hierarchy of Star Formation}

Hierarchal structure in the ISM presumably comes from self-gravity and
turbulence. This gas structure continues to sub-stellar scales, as
shown by high resolution molecular observations. The densest regions,
which are where individual stars form, are often clustered into the
next-denser regions. Stars form in the densest regions, at first
somewhat independently it seems, and then they move around, possibly
interact, and ultimately mix together inside the next-lower density
region. That mixture is the ``cluster.'' More and more subclusters mix
over time until the cloud disrupts. The net efficiency of star
formation (fractional star mass) is automatically high on small and
dense scales because of the hierarchy of structures.

Clusters are the inner mixed parts of the hierarchy of young stellar
structures. The hierarchy that is present on larger scales was also
present on smaller scales before this mixing. If we consider a
cluster's ``Nature'' to be its internal properties, and its ``Nurture''
to be its external properties, then Nature and Nurture for a star
cluster are essentially the same thing. Internal and external depend on
the extent to which gravitational mixing has occurred at the time of
observation.

\section{Unclustered Star Formation}

Barba et al. (2009) studied the giant star-forming region NGC 604 in
M33 with HST NICMOS to look for obscured young clusters. They found
that it contains almost no clusters but that most star formation is
distributed, in agreement with Hunter et al. (1996).  Giant molecular
clouds also show some star formation in a distributed form (Megeath et
al. 2004; J{\o}rgensen et al. 2006, 2007), although in the solar
neighborhood, most star formation is clustered. Ma\'iz-Apell\'aniz
(2001) generalized the discussion by considering three types of
clustering: 1. compact clusters with weak unclustered halos typically
measuring 50x50 pc$^2$; 2. compact clusters with strong halos,
measuring 100x100 pc$^2$, and 3. purely hierarchical star formation
with no clusters.

A possible reason for variations in young stellar clustering is
explored by Elmegreen (2008). The basic point was outlined above, that
hierarchical star formation automatically leads to high efficiencies
and bound stellar clustering at high densities. The point now is to
find the mass fraction of star formation that has a sufficiently high
efficiency to produce a bound cluster after the gas leaves. This
fraction can be calculated from the density pdf in a GMC. The density
pdf has a log-normal form in isothermal turbulence and a power-law form
at high density when self-gravity is important (Klessen 2001). An
integral over density times the density pdf gives the mass. The
integral illustrates the importance of the breadth of the density pdf:
for a narrow pdf (i.e., with a low dispersion), the critical efficiency
is reached at a high density and a low mass fraction, while for a wide
pdf, the critical efficiency is reached at a low density and a high
mass fraction. What matters is the slope of the pdf at the density
where the efficiency becomes critical for producing a bound cluster.
Thus the breadth of the pdf enters into this slope. Also entering is
the centroid density. Generally the pdf is in the high-density falling
part at the critical value for high efficiency. Then a shift in the pdf
toward higher density will also make the slope shallower there (for the
log-normal case). Thus either a broader pdf or a higher average density
will lead to a higher fraction of the star-formation mass in the form
of bound clusters.  A broader pdf could arise from a higher turbulent
Mach number. It follows that higher pressure regions are more likely to
form bound clusters.

\section{Cluster Mass Functions}

The mass function for bound clusters is approximately $n(M)dM\propto
M^{-2}dM$ for a given age range (Elmegreen \& Efremov 1997; Zhang \&
Fall 1999; de Grijs et al. 2003; de Grijs \& Anders 2006). This
function is also observed for various ages, although the lower limit to
the power law depends on the age because of cluster magnitude detection
limits and cluster fading with age (Elmegreen \& Efremov 1997; Hunter
et al. 2003; de Grijs \& Anders 2006).

The cluster mass function resembles the mass function of
cluster-forming cloud clumps (Reid \& Wilson 2005; Rathborne et al.
2006), and both presumably get their form from the hierarchical
structure of interstellar matter.  In a hierarchy, all of the mass is
represented at all levels, so the total mass in each level (which has
logarithmic intervals of mass), $Mn(\log M)d\log M$, is the same as the
total mass in any other level. It follows that $Mn(\log M)d\log M=$
constant$\times d\log M$, $n(\log M)\propto M^{-1}$, and $n(M)\propto
M^{-2}$. Similarly if sub-clouds are randomly selected from a
hierarchical cloud, then their distribution function will be $M^{-2}$
too. Or, if we generate a fractal Brownian motion cloud, the mass
distribution of clumps will be close to $M^{-2}$, with a preferred
value of the power spectrum given by the Kolmogorov turbulence law
(Elmegreen et al. 2006).

Recent observations suggest that the power law mass function sometimes
has an upper mass cutoff, making it a Schechter function (as originally
applied to galaxies -- Schechter 1976).  These cutoffs have been
observed in M51 (Gieles et al. 2006ab) and several other galaxies
(Waters et al. 2006; Larsen 2009).

Before we turn to halo globular clusters, the essential points of the
preceding sections can be summaries as follows:
\begin{itemize}
\item Gas is hierarchical in space and time, presumably as a result of
turbulence and self-gravity.
\item Therefore star formation is hierarchical.
\item Therefore the efficiency of star formation ($M_{\rm stars}/M_{\rm total}$)
increases with average density.
\item Therefore bound objects (which require high efficiency) form at high
density. The result is a cluster.
\item Finally, the mass function of all this structure is a power law with a slope
$-2$ for equal intervals of mass.
\item The origin of the upper cut off mass for clusters is unexplained
\end{itemize}

\section{Globular Clusters}

The mass functions for galactic clusters, reviewed above, resemble the
upper parts of the mass functions for globular clusters, which appear
to have lost their low mass members. Jordan et al. (2007) fit power
laws to the upper parts of the globular cluster mass functions in Virgo
galaxies ($3-20\times10^5\;M_\odot$ clusters). They found that the
upper mass power law is steeper for lower luminosity galaxies, and the
width of the log-normal mass function is also smaller for lower
luminosity galaxies.  This result implies that low-mass galaxies have a
smaller upper mass cutoff for their clusters. Waters et al. (2006)
measured the mass function for M87 globular clusters with HST and also
found that the best fit was an evolved Schechter function with a cutoff
mass of $M_c=10^{6.4}$. The evolution models in Waters et al. assumed
various mass dependencies for cluster evaporation, with the best fit
having no mass dependence. For either this standard model of mass loss
rate, $dM/dt=$const, or the Lamers et al. (2006) model, $dM/dt\propto
M^{0.38}$, the low mass globular clusters evaporate first, leaving only
the high mass clusters after a Hubble time.

The problem with evaporation models is that globular cluster mass
functions are independent of galactocentric radius in galaxies (Kundu
et al. 1999; Jordan et al. 2007), whereas the evaporation rate depends
on the tidal density, which depends on the galactocentric radius. Thus
the outer regions of galaxies should have more low-mass globular
clusters remaining than the inner regions, but they do not. One
proposed solution is that globular cluster orbits are highly radial
(Fall \& Zhang 2001), but this disagrees with the observed profile of
globular cluster velocity dispersion versus radius (Vesperini et al.
2003).

McLaughlin \& Fall (2008) separated the Milky Way globular cluster
luminosity function into three groups according to the density at
half-light radius. The peak mass was found to depend on the density as
expected for $dM/dt=$constant evaporation of a Schechter function, and
the mass functions are independent of position, as required. Chandar et
al. (2007) found the same thing for M104. However, low-density globular
clusters are lower mass anyway because they all have about the same
radius (e.g., McLaughlin 2000). Perhaps density-dependent evaporation
is the reason they all have the same radius, but this has not been
demonstrated independently.

Another option is that the globular cluster mass function was peaked
from a young age (Vesperini 2000; Parmentier \& Gilmore 2007) and has
not evolved much since then. Evaporation tends to preserve a peak in
the mass function once it forms (Vesperini 1998). Parmentier, et al.
(2008) showed that cluster disruption during initial gas removal can
convert a power law mass function into a log-normal mass function at a
very early stage. This would satisfy the observations that show no
radial gradients in the mass function. The problem with this model is
that modern clusters do not lose their initial power law mass functions
so quickly. The difference could be that halo globular clusters formed
with a lower efficiency ($<25$\%) than disk clusters ($\sim40$\%)
because cluster disruption is more likely during gas loss for low
efficiencies.  We note that in the hierarchical model of cluster
formation, lower efficiency for a given cluster density (required to
define an object as a cluster) follows from lower ISM pressures
(Elmegreen 2008), and that is consistent with globular cluster
formation in dwarf galaxies.

Halo globular clusters differ from disk open clusters in several ways.
Globulars are more massive and this presumably allows self-enrichment
of heavy elements from red giant winds. Globular clusters are also
older, and this means that  evolutionary effects are more prominent,
such as evaporation and core collapse. Globular clusters are lower in
metallicity than disk clusters, and this means that stellar winds are
weaker and any self-enrichment is more obvious. Globular cluster are
also in the halo and many appear to come from dissolved companion dwarf
galaxies. Others could come from the starburst phase of a merger when
the disk and its clusters are scattering into the halo.  Yet others
could come from star formation in small galaxies before the large
galaxies formed.

Harris (2009) used HST to observe six giant elliptical galaxies within
40 Mpc.  He found 7800 globular clusters with half-light radii larger
than 1.5 pc. The blue (low metallicity) and red (high metallicity)
sequences found previously for many globular clusters systems are clear
in this new data. The blue sequence turns slightly redder at high mass,
suggesting self-enrichment for $M>10^6\;M_\odot$. Normal blue globulars
have [Fe/H]$\sim-1.5$, while in his survey, the most massive blue
globulars have [Fe/H]$\sim-1$. This redward trend toward higher mass
among the blue globulars was also found in M87 by Peng (2009). Bailin
\& Harris (2009) suggest supernova ejecta is trapped in the more
massive clusters during the gas-rich formation phase. Similar
observations are in Harris et al (2006), Strader et al. (2006), and
Mieske et  al. (2006).

Bedin et al. (2004) found multiple main sequences and main sequence
turnoffs in $\omega$Cen. The multiple main sequence implies there is a
range of He abundances (D'Antona et al. 2002; Norris 2004), while the
multiple subgiant branches mean there is a range of ages (Milone et al.
2008) or CNO abundances (Cassisi et al. 2008).  Self-enrichment is
likely.  Ventura et al. (2009) found a split subgiant branch in NGC
1851. They modeled the second generation with a CNO abundance three
times that of the first generation. They suggested that massive AGB
stars in the first generation have ejecta with five times the CNO
abundances of the stars themselves, and this ejecta was diluted by 50\%
with pristine gas to keep the He abundance low.  CNO enrichment is
usually so large in globular clusters that the progenitor stars have to
outnumber the cluster stars. Either the former cluster was much more
massive (D$^\prime$Ercole et al. 2008) or the globular cluster was the
core of a dwarf galaxy which collected ejecta from many other clusters
(Bekki \& Norris 2006).

Marcolini (2007) did hydrodynamic models of star formation and metal
production in a dwarf Spheroidal galaxy with an $\omega$ Cen-type
cluster forming in the nucleus.  They assumed star formation had a high
rate for 1 Gyr and then a lower rate for another 0.6 Gyr. The
$\alpha$-elements form from SNII quickly and Fe forms from SNIa more
slowly. The SNII push out and mix with the nuclear gas, producing a
uniform and growing Fe abundance. This gas resettles to the center when
the massive stars are gone. SNIa then become important, and they
pollute the central region locally, leading to pockets of low
$\alpha$/Fe. The result is a wide range of [Fe/H] in globular cluster
stars, with high $\alpha/Fe$ ratios at low [Fe/H], and an increasing
dispersion toward low $\alpha/Fe$ at high [Fe/H] abundances, as
observed.

Another possibility is that small clusters merge in the nucleus of a
dwarf galaxy. Georgiev et al. (2009) found two globular clusters in the
dwarf Irr UGCA and thought they should merge in 0.4 Gyr by dynamical
friction and make a nuclear globular cluster.

The Sagittarius dwarf galaxy is a good example of a source for Milky
Way globular clusters, having Terzan 7, Terzan 8, Arp 2, and M54 as
part of its orbital debris.  Carraro et al. (2007) suggested that the
globular cluster Whiting 1 is also likely associated with the Sgr
dwarf, which would make 6 total. The age of Whiting 1 is $6.5\pm0.7$
Gyr an [Fe/H]$= -0.4$ to $-1.1$, which are consistent with the
age-metallicity pattern in the Sgr dwarf. Whiting 1 also has an
extended luminosity profile, presumably a tidal tail, and its position,
distance, and velocity place it in the Sgr stream.

Casetti-Dinescu et al. (2009) considered the Virgo Stellar Stream and
integrated its stellar orbits back for 5 Gyr. They found that the
globular cluster NGC 2419 could be in this stream. Gao et al. (2007)
found common streams for many Milky Way globular clusters, based on
common energy, angular momentum and orbital poles. Gao et al. suggest
that 20\% of globular clusters are in common streams.

Smith et al. (2009) noted that four globular clusters lie in a halo
star kinematic overdensity. Their metallicities are [Fe/H]$=-2.22$,
$-1.54$, $-1.58$ and $-1.65$; one of them, NGC5466, is disrupting
(Odenkirchen \& Grebel 2004; Belokurov et al. 2006). Other related
studies are in Dinescu et al. (1999), Palma et al. (2002); and Mackey
\& Gilmore (2004).

\section{Summary}

Stars form from gas in hierarchical patterns. Clusters are the inner
mixed part of this star formation distribution. In such a hierarchy,
the star formation efficiency is automatically high on small scales,
and if it is high enough, the cluster remains bound after the gas
leaves. The cluster mass function, $dN/dM\sim M^{-2}$, follows from the
hierarchy too. There is a possible upper cutoff mass, $M_c$ making the
mass function something like $M^{-2}e^{-M/Mc}$. Halo globular clusters
may have a similar upper mass cutoff and they all drop toward lower
mass, without the power law. It is not yet understood if the missing
low-mass clusters were removed by a long-term process of evaporation,
or if they were destroyed early by some other process, including gas
loss after star formation. Globular clusters often show multiple star
formation events, which implies self-enrichment over a period of 0.1 to
1 Gyr. Perhaps their high mass and low metallicity is enough to produce
this enrichment, which is not generally observed for low mass clusters
today. Or perhaps globular clusters were dwarf galaxy nuclei that
collected independent clusters with a wide range of metallicities. Many
Milky Way globular clusters orbit in the stellar debris streams from
former dwarf galaxy companions.

\end{document}